\documentclass[twocolumn,letterpaper,aps,prl,
superscriptaddress,showpacs,amsmath]{revtex4-1}

\usepackage{dcolumn,graphicx,color,booktabs,microtype,afterpage} \graphicspath{{Figures/}}

\usepackage{graphicx}
\usepackage{dcolumn}
\usepackage{bm}
\usepackage{amsmath}
\usepackage{color}

\def\seven{Sr\ensuremath{_3}Fe\ensuremath{_2}O\ensuremath{_{7}}}
\def\TN{\ensuremath{T_\text{N}}}

\begin{document}
\preprint{}

\title{Competing exchange interactions on the verge of a metal-insulator
  transition in the two-dimensional spiral magnet \seven}

\author{J.-H. Kim}
\affiliation{Max-Planck-Institut f\"ur Festk\"orperforschung, D-70569
  Stuttgart, Germany}
\author{Anil Jain}
\affiliation{Max-Planck-Institut f\"ur Festk\"orperforschung, D-70569
  Stuttgart, Germany}
\affiliation{Solid State Physics Division, Bhabha Atomic Research Centre, Mumbai 400085, India}
\author{M.~Reehuis}
\affiliation{Helmholtz-Zentrum Berlin f\"ur Materialien und Energie, D-14109
  Berlin, Germany}
\author{G.~Khaliullin}
\affiliation{Max-Planck-Institut f\"ur Festk\"orperforschung, D-70569
  Stuttgart, Germany}
\author{D.~C.~Peets}
\affiliation{Max-Planck-Institut f\"ur Festk\"orperforschung, D-70569
  Stuttgart, Germany}
\author{C.~Ulrich}
\affiliation{Max-Planck-Institut f\"ur Festk\"orperforschung, D-70569
  Stuttgart, Germany}
\affiliation{School of Physics, University of New South Wales, Sydney, NSW 2052, Australia}
\affiliation{Australian Nuclear Science and Technology Organisation, Lucas Heights, NSW 2234, Australia}
\author{J.~T.~Park}
\affiliation{Forschungsneutronenquelle Heinz Maier-Leibnitz (FRM-II), D-85748
  Garching, Germany}
\author{E.~Faulhaber}
\affiliation{Forschungsneutronenquelle Heinz Maier-Leibnitz (FRM-II), D-85748
  Garching, Germany}
\author{A.~Hoser}
\affiliation{Helmholtz-Zentrum Berlin f\"ur Materialien und Energie, D-14109
  Berlin, Germany}
\author{H.~C.~Walker}
\affiliation{ISIS Facility, STFC, Rutherford Appleton Laboratory, Chilton, Didcot, Oxfordshire, OX11-0QX, United Kingdom}
\author{D.~T.~Adroja}
\affiliation{ISIS Facility, STFC, Rutherford Appleton Laboratory, Chilton, Didcot, Oxfordshire, OX11-0QX, United Kingdom}
\affiliation{Physics Department, Highly Correlated Matter Research Group, University of Johannesburg, PO Box 524, Auckland Park 2006, South Africa}
\author{A.~C.~Walters}
\affiliation{Max-Planck-Institut f\"ur Festk\"orperforschung, D-70569
  Stuttgart, Germany}
\author{D.~S.~Inosov}
\affiliation{Max-Planck-Institut f\"ur Festk\"orperforschung, D-70569
  Stuttgart, Germany}
\affiliation{Institut f\"ur Festk\"orperphysik, TU Dresden, D-01069 Dresden, Germany}
\author{A.~Maljuk}
\affiliation{Max-Planck-Institut f\"ur Festk\"orperforschung, D-70569
  Stuttgart, Germany}
\affiliation{Leibniz Institut f\"ur Festk\"orper- und Werkstoffforschung,
  D-01171 Dresden, Germany}
\author{B.~Keimer}
\email{b.keimer@fkf.mpg.de}
\affiliation{Max-Planck-Institut f\"ur Festk\"orperforschung, D-70569
  Stuttgart, Germany}

\date{\today}

\begin{abstract}

We report a neutron scattering study of the magnetic order and dynamics of the
bilayer perovskite \seven, which exhibits a temperature-driven metal-insulator
transition at 340~K. We show that the Fe$^{4+}$ moments adopt incommensurate
spiral order below $T_\text{N}=115$~K and provide a comprehensive description of the
corresponding spin wave excitations. The observed magnetic order and
excitation spectra can be well understood in terms of an effective spin
Hamiltonian with interactions ranging up to third nearest-neighbor pairs.
The results indicate that the helical magnetism in \seven\ results from
competition between ferromagnetic double-exchange and antiferromagnetic
superexchange
interactions whose strengths become comparable near the metal-insulator
transition. They thus confirm a decades-old theoretical prediction and provide
a firm experimental basis for models of magnetic correlations in strongly
correlated metals.

\end{abstract}


\maketitle

Following theoretical progress including the development of the dynamical
mean-field theory~\cite{Geo97}, the description of correlation-driven
metal-insulator transitions (MITs) has recently been rapidly advancing, but
realistic calculations of magnetic correlations near MITs remain a formidable
challenge. Manganese oxides have served as prominent model materials for
research on magnetism in proximity to MITs.
Whereas localized spins in insulating manganates interact via (predominantly
antiferromagnetic) superexchange interactions mediated by high-energy virtual
excitations of the Mn $d$-electrons, ferromagnetic double-exchange
interactions mediated by itinerant electrons dominate in their metallic
counterparts. In terms of dynamical mean field theory, superexchange and
double-exchange interactions originate in high-energy incoherent
(``Hubbard-like'') and narrow quasiparticle (``Kondo-like'') bands,
respectively~\cite{Kha05}. According to a long-standing
prediction~\cite{deG60}, competition between these antagonistic exchange
interactions generates non-collinear magnetic structures in the vicinity of
MITs. Unfortunately, disorder-induced electronic phase separation near
MITs~\cite{Dagotto} has thus far largely precluded experimental tests of this
prediction in manganates and other transition metal oxides (TMOs).


Non-collinear magnetism has been observed in a small number of disorder-free
model materials (including NdNiO$_3$~\cite{Scagnoli} and
CaFeO$_3$~\cite{Woodward}) that exhibit temperature-driven MITs, and in TMO
superlattices where MITs can be driven by adjusting the doping
level~\cite{Santos} or electronic dimensionality~\cite{Frano} without
introducing
disorder. Following the original prediction~\cite{deG60}, these structures
have been discussed on a qualitative level in terms of competing superexchange
and double-exchange interactions~\cite{Kha05,Santos,Balents}, but alternative
interpretations have also been proposed. In particular, it was pointed out
that double exchange alone can generate spiral magnetism in TMOs with metal
ions in high oxidation states, where the usual charge transfer between metal
$d$- and oxygen $p$-states is at least partially
reversed~\cite{Mos05,Li}. Both models differ in their predictions for the
magnitudes and spatial range of the exchange coupling between electron spins
on different lattice sites. In principle, these parameters can be extracted
from the spin wave dispersions measured by inelastic magnetic neutron
scattering (INS). To the best of our knowledge, however, such measurements have not
been reported for stoichiometric model materials, because crystals of
sufficient size and quality have not been available.

In order to guide and test theoretical concepts of magnetic order near MITs,
we have used neutron scattering to investigate the magnetic structure and
dynamics of fully oxygenated \seven, a stoichiometric compound built up of bilayers of FeO$_6$ octahedra
that was recently found to undergo a continuous transition from a
high-temperature metallic to a low-temperature insulating phase at temperature
$T_\text{MIT}=340$~K~\cite{Peets2013}. This material is based on Fe$^{4+}$ ions
(nominal electron configuration 3$d^4$), which are isoelectronic to Mn$^{3+}$ (3$d^4$)
and closely analogous to Ru$^{4+}$ (4$d^4$) ions in two extensively studied
families of manganates and ruthenates, and it is  isostructural to the
compounds La$_{2-x}$Sr$_x$Mn$_2$O$_7$ and Sr$_3$Ru$_2$O$_7$ that are well
known for their magnetoresistive properties~\cite{Tokura} and electronic
liquid-crystal behavior~\cite{Borzi}, respectively. Despite these analogies,
we find that below the N\'eel temperature, $T_\text{N} = 115$~K, \seven\ exhibits a spiral state that has
no analog in manganates or ruthenates. Unlike other model
materials~\cite{Scagnoli,Woodward}, single crystals with quality and volume
sufficient for INS measurements have been
grown~\cite{Peets2013,Maljuk2004}. We have thus been able to determine the
spin wave dispersions and extract the exchange parameters, which compare
favorably with models based on competing superexchange and double-exchange
interactions~\cite{Kha05}. In contrast, predictions based on double exchange
alone~\cite{Mos05} do no yield satisfactory agreement with the data. Our data
thus quantitatively confirm the long-standing prediction of competing exchange
interactions near MITs~\cite{deG60}, and they establish \seven\ as a
two-dimensional model material for spiral magnetism, which has recently
attracted considerable attention in the context of research on
multiferroicity, topological excitations~\cite{Nagaosa}, and copper- and
iron-based high-temperature superconductors~\cite{Lindgard,Sushkov,Stock}.

\begin{figure}
\begin{center}
\includegraphics[width=\columnwidth]{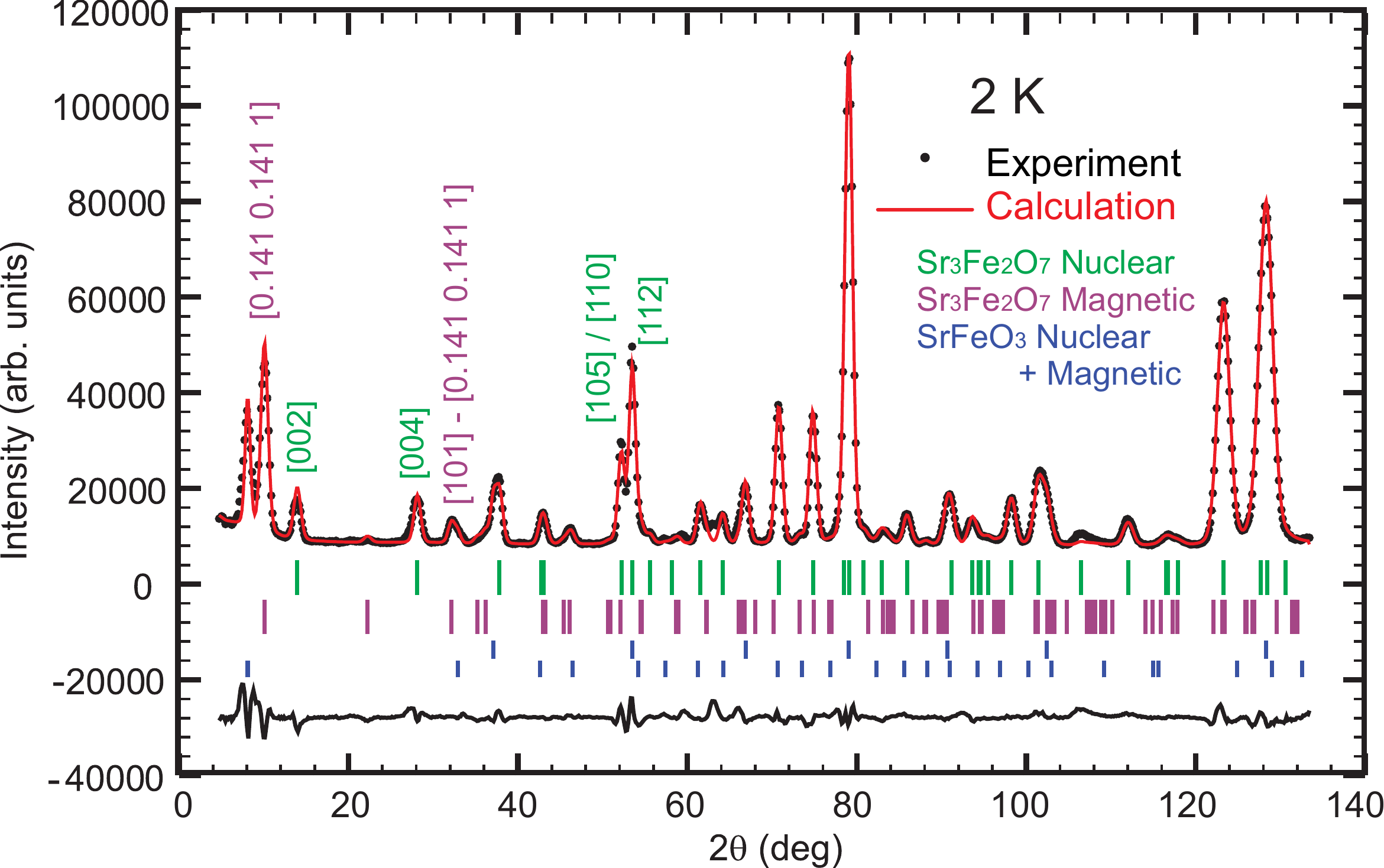}
\caption{\label{fig:mag} (Color online) Neutron powder diffraction data of \seven\ at 2
  K. The helical spin modulation vector in \seven\ was found to be $\vec{k} =
  [0.141~0.141~1]$ (in the tetragonal setting). The upper symbols and
  line represent experimental data and calculated results, respectively. A small amount of SrFeO$_3$ (\TN = 133~K)
in our powder sample manifests its helical magnetic phase with $\vec{k} =
[0.129, 0.129, 0.129]$, consistent with fully-oxygenated
SrFeO$_3$~\cite{Reehuis2012}.
  Upper, middle, and bottom bars indicate the calculated positions of
  \seven\ nuclear, \seven\ magnetic, and SrFeO$_3$ reflections, respectively,
  and the bottom line is a residual.}
\end{center}
\end{figure}

The experiments were performed on a powder sample of weight 3~g, and on
cylindrical single crystals of diameter 5~mm and length 5~mm (5~cm) for
elastic (inelastic) scattering. Their preparation and
characterization are described in Refs.~\onlinecite{Maljuk2004,Peets2013}.
All samples were characterized by neutron and x-ray diffraction,
as described in Ref.~\onlinecite{Peets2013}. The magnetic structure
was determined from single-crystal neutron diffraction data obtained on the E5
diffractometer (neutron wavelength $\lambda = 2.38$~\AA), and powder data from the E6 diffractometer ($\lambda = 2.44$~\AA) at BER-II
(Helmholtz-Zentrum Berlin, Germany), respectively. INS experiments on a single
crystal were carried out on neutron triple axis spectrometers (TAS)
PANDA (cold TAS) and PUMA  (thermal TAS) at the FRM-II (Munich, Germany) and direct-geometry time-of-flight (TOF) spectrometer MERLIN at the ISIS spallation neutron source (Didcot, UK). At PANDA, the momentum of the scattered neutrons was set to $k_\text{f}= 1.57$ \AA$^{-1}$, and  at PUMA, $k_\text{f}= 2.662$ \AA$^{-1}$ was used. For the TOF measurements, the incident energy was 60~meV  and the sample was mounted with the $(H\kern.5pt\bar{H}\kern.5ptL)$ scattering plane horizontal. The magnetic excitations throughout the Brillouin Zone in all
symmetry directions were mapped out by  rotating the crystal over $60^{\circ}$ in steps of $1^{\circ}$ about the vertical $[110]$ axis, starting from $k_\text{i}$ along $[1\bar{1}0]$.  The TOF data were transformed into units of energy and momentum transfer using the Horace software~\cite{horace}.


\begin{figure}
\begin{center}
\includegraphics[width=\hsize]{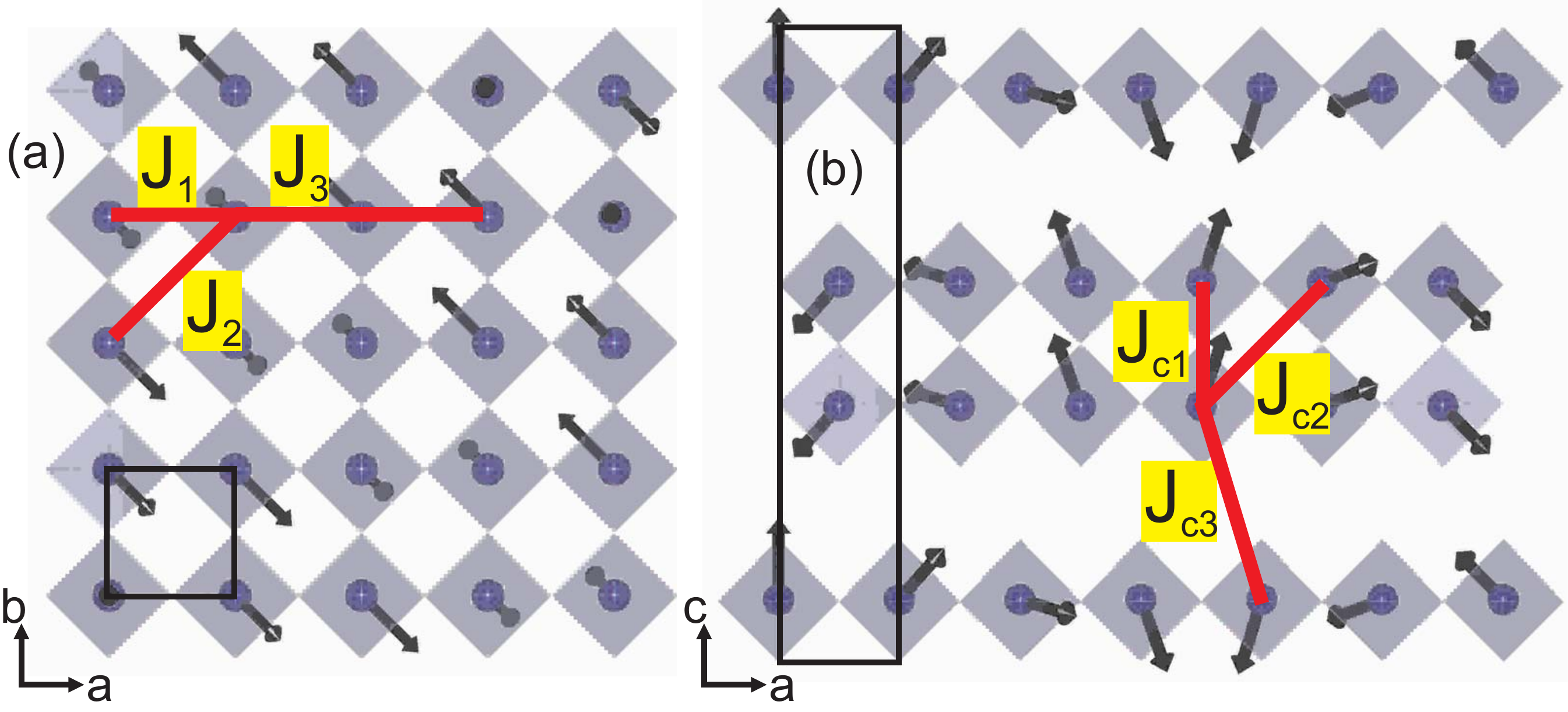}
\caption{(Color online) Helical magnetic structure of \seven\ projected onto the (a) $ab$
  and (b) $ac$ planes. The rectangle indicates the tetragonal unit cell.
  Spheres and octahedra represent the Fe ions and FeO$_6$ units,
  respectively. Arrows indicate the spin directions. Lines indicate
the spin exchange couplings $J$ included in the theoretical fits.}
\label{fig:magstr}
\end{center}
\end{figure}

Figure~\ref{fig:mag} shows neutron powder diffraction data on \seven\ at $T=2$
K (well below $T_N$).
They were refined using FullProf~\cite{Fullprof} based on the published
tetragonal space group $I4/mmm$ (No. 139) with lattice parameters
$a=b=3.846(4)$ \AA~and $c= 20.234(2)$ \AA~(at $T=390$~K). The best fit, with
$R_F = \frac{ \sum||F_{obs}|-|F_{calc}||}{\sum|F_{obs}|} = 0.061$, implies
that
\seven\ has an incommensurate magnetic propagation vector of $\vec{k} =
[\xi~\xi~1]$ with $\xi$ = 0.1416(3). In addition to the powder refinements,
single-crystal neutron diffraction data measured at $T=10$ K were refined with
$R_\text{B} = \frac{\sum ||I_\text{obs}| -|I_\text{calc}||}{\sum|I_\text{obs}|}=0.093$, based on 142
magnetic reflections (26 unique). The refinement results of the powder (2~K)
and single crystal (10~K) data are almost identical. The magnetic
structure of \seven\ obtained in this way is helical with an elliptical helix
having a $c$-component significantly smaller than its $ab$-components
[powder : $\mu_{a,b} = 3.53(4) \mu_{B}$ and $\mu_{c} = 3.04(5) \mu_{B}$,
single crystal: $\mu_{a,b} =3.58(11) \mu_{B}$ and $\mu_{c}\ = 3.19(5) \mu_{B}$].
As illustrated in Fig.~\ref{fig:magstr}, all spins lie in a plane
perpendicular to the [1~1~0] direction. Along the $c$-axis, the spins of iron
atoms at $(0~0~\pm z)$ are anti-parallel with those at $(0.5~0.5~0.5\pm z)$
[see Fig.~\ref{fig:magstr}(a)], which yields the $c$-component of the
propagation vector $k_z=1$. This helical structure is analogous to those
in metallic SrFeO$_3$ and insulating CaFeO$_3$ which have a rotation axis of
[1~1~1]~\cite{Tak72,Woodward,Reehuis2012}.


\begin{figure*}[t]\vspace{-1.2em}
\includegraphics[width=\textwidth]{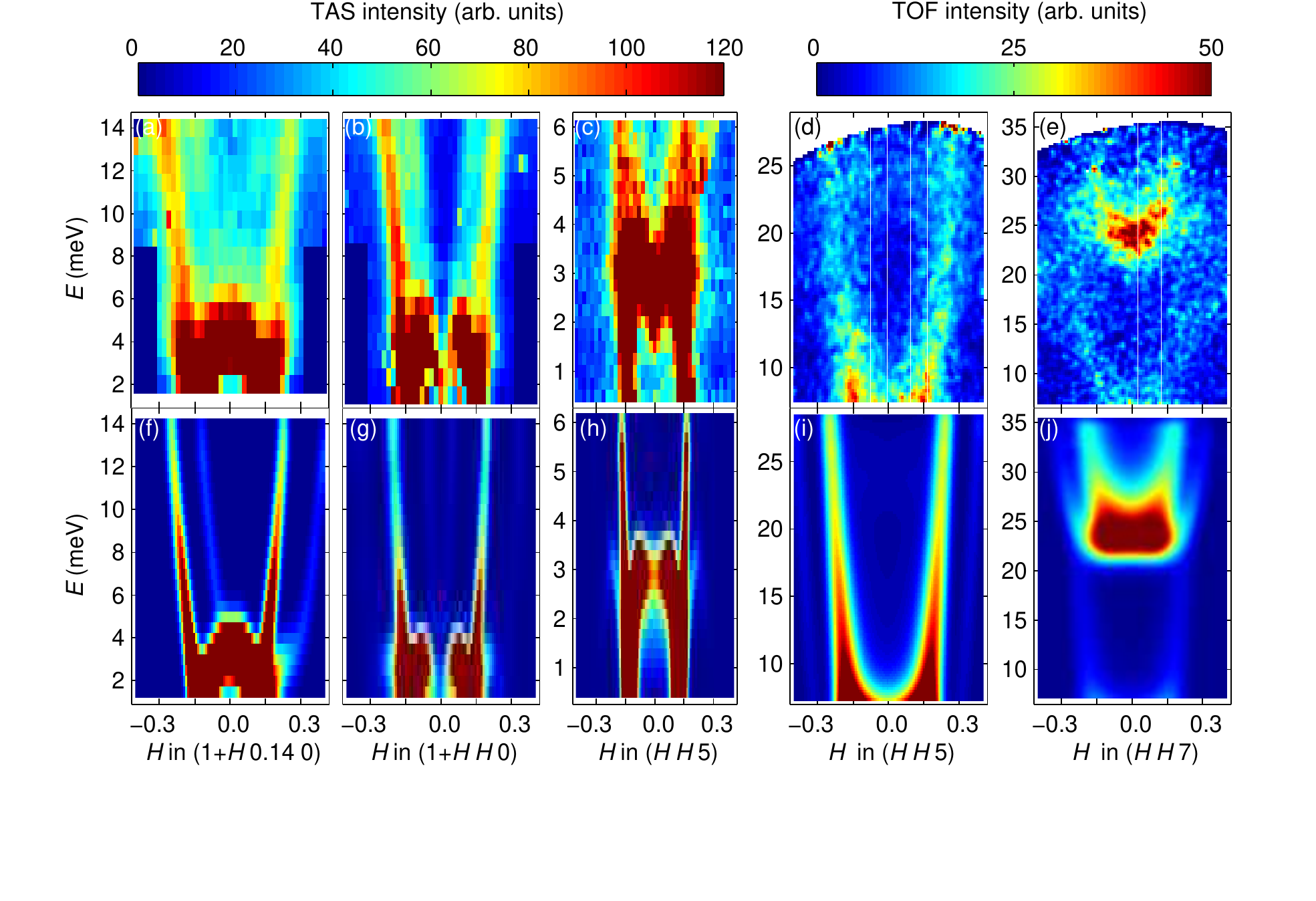}
\vspace{-6em}
\caption{(Color online) Contour maps of spin-wave dispersions in \seven~at 7~K along
   the (a)~$[H\kern.5pt0\kern.5pt0]$, (b)~$[H\kern.5ptH\kern.5pt0]$, (c-d)~$[H\kern.5ptH\kern.5pt5]$, and (e)~$[H\kern.5ptH\kern.5pt7]$ directions. The intensities in (a) and (b) were multiplied by $\sqrt{E}$ to enhance the upper band. (f)-(j) show the
  theoretical dispersions and neutron scattering intensities (convoluted with the instrumental resolution), calculated with the parameters described in the text.}
\label{fig:thermal}
\end{figure*}

In order to study the dynamical magnetic properties of \seven, we performed
INS measurements on a single crystal at $T=4$~K, well below $T_\text{N}$. Figures~\ref{fig:thermal}~(a) and (b) show contour maps of neutron
scattering intensities obtained from PUMA up to 14 meV along the
$[H00]$~and $[H\kern.5ptH\kern.5pt0]$~directions, respectively. To quantify the lower-energy excitations
more precisely, the spin excitations along the $[H\kern.5ptH\kern.5pt5]$ direction [Fig.~\ref{fig:thermal}~(c)] were measured at the PANDA cold-neutron
TAS, where no significant anisotropy gap  is observed. In a bilayer system, one typically expects two spin wave branches, optic and acoustic, as observed in high-temperature superconducting cuprates~\cite{Tranquada} and bilayer manganates~\cite{Perring}. To detect the optical branch, high-energy INS measurements were performed on the MERLIN TOF spectrometer, and the results are summarized in Figs.~\ref{fig:thermal}~(d) and (e). We were able to separate the two spin wave branches due to the different dependencies of their INS cross section on the \emph{L} (\emph{c}-axis) component of the momentum transfer $\vec{q}$. The optic (acoustic) branch has maximum cross section when $L \Delta z_\text{Fe}$ is half-integer (integer). Here $\Delta z_\text{Fe}= 0.195$ is the distance between nearest-neighbor Fe spins within one bilayer, expressed as a fraction of the lattice constant \emph{c}. The INS signal at $L = 5$, shown in Fig.~\ref{fig:thermal}~(d),  originates from the acoustic branch. However, the observed strong INS signal, for $\sim$~22~meV and higher energies [Fig.~\ref{fig:thermal}~(e)], at $L = 7$, mostly comes from the optic branch. The optic spin-wave energy gap allowed us to determine the intra-bilayer coupling ($J_\text{c1}$ in Fig.~\ref{fig:magstr}).

As far as low-energy magnon dispersions are concerned, double-exchange
systems can be described by spin-only models~\cite{Kha00}. Thus, the above
physical considerations can be cast into the following phenomenological
spin Hamiltonian:
\begin{equation}
H=\sum_{ij} J_{ij} \mathbf{S}_i \cdot \mathbf{S}_j
+\Delta \sum_{i}S_{i\alpha}^2 ,
\label{eq:Hamiltonian}
\end{equation}
where $J_{ij}$ represents the Heisenberg coupling between the $i$-th and
$j$-th spins $\mathbf{S}$, $\Delta>0$ is the easy (110) plane anisotropy
parameter, and $S_\alpha$ refers to the spin component along the [110] axis.

We have carried out standard linear spin-wave calculations~\cite{Toth}, using a minimal
set of input parameters in Eq.~(\ref{eq:Hamiltonian}), to fit the magnon
dispersion and intensities observed. In order to describe the main features
of the experimental data on a quantitative level, the minimal spin
Hamiltonian is found to
contain five different exchange couplings. These are the nearest-neighbor
$J_1$ and longer-range $J_2, J_3$ interactions between Fe-spins within the
$ab$-plane [see Fig.~\ref{fig:magstr}(a)], and the $c$-axis couplings
$J_\text{c1}$ and $J_\text{c3}$ that stand for the intra- and inter-bilayer
interactions, respectively [see Fig.~\ref{fig:magstr}(b)]. We have assumed
that equivalent domains of the spiral structure
contribute equally to the spectra, and convoluted the calculated intensities
with the instrumental resolution function. Figures~\ref{fig:thermal}(f)-(j)
present the results of the calculations with the following parameters:
$J_{1} = -7.2$ meV, $J_{2} = 1.05$ meV, $J_{3} = 2.1$ meV, $J_\textup{c1}=-5.1$ meV,
$J_\textup{c2} < 0.01$ meV, $J_\textup{c3} = 0.01$ meV, $\Delta = 0.06$ meV. Here positive (negative)
values of $J$ correspond to antiferromagnetic (ferromagnetic) interactions.
The calculated spin-wave dispersions show a remarkable agreement with the experimental data.


We now turn to the interpretation of our results. The helical
order observed and the overall topology of the magnon dispersions are
consistent with
the negative charge-transfer energy model~\cite{Mos05}. However, the predicted
extremely soft dispersion ($\sim 10^{-3}$ of the full magnon bandwidth) of
spin-waves for $\vec{q}$ values below the ordering vector $\vec k$ is
inconsistent with the data. As shown in Figs.~\ref{fig:thermal}(a)-(c), we observe highly dispersive magnons in the window
$\vec{q}<\vec k$, reaching a maximum of $\sim 3$~meV. This is more than
an order of magnitude larger than the calculated value of $\sim 0.1$~meV for a
three-dimensional model~\cite{Mos05}.

In a scenario based on competing ferromagnetic double-exchange and
antiferromagnetic superexchange interactions,
the balance between both couplings is controlled by the degree of
the itineracy/localization of the conduction electrons. The signs and relative
strengths of the exchange interactions are consistent with theoretical
estimates~\cite{Kha05}. Specifically, the strength of
the double-exchange coupling $J_\text{DE} \simeq -\frac{1}{4S^2}\tilde{t}$, where
$\tilde{t}=\kappa t$ with $\kappa<1$ is the effective hopping amplitude.
Enhanced correlations near the metal-insulator transition may considerably
reduce the electron mobility and hence the values of $\kappa$ and $J_\text{DE}$.
In particular, the choice of $\kappa \sim 0.3$ provides $J_\text{DE} \sim -10$~meV
which is comparable to typical values of antiferromagnetic superexchange
interactions in manganates,
$J_\text{SE}\sim\frac{1}{4S^2}\frac{t^2}{U}\sim 3$~meV. Whereas spin wave
measurements on isostructural manganates~\cite{Perring,Hirota} are well
described by nearest-neighbor exchange Hamiltonians, the longer-range
interactions observed in \seven\ are consistent with the larger $p-d$
covalency predicted in theoretical work on the iron
oxides~\cite{Kha05,Mos05,Li}.

A few comments are in order regarding the $J$ values we obtained from the
above fits. The large negative values of the nearest-neighbor couplings
$J_{1}, J_\text{c1}$  imply that ferromagnetic
double-exchange in ferrates is sufficiently strong to keep
the neighboring spins roughly parallel. (Note that $J_1\sim -|J_\text{DE}|+J_\text{SE}$.)
However, the longer-range antiferromagnetic exchange interactions
$J_2, J_3$ are sizable, and in fact, they are
crucial to stabilize the spin-helix against simple ferromagnetic and
canted-antiferromagnetic states~\cite{Sha02}. A total-energy calculation as a function of wave vector shows that these exchange parameters are compatible with the observed helical spin structure.

Remarkably, the coupling constant $J_3$ is larger than the second-neighbor
interaction $J_2$.
This can be understood as a consequence of the large $pd\sigma$ virtual
charge-transfer along the Fe-O-Fe-O-Fe line.
We also note that the $ab$-plane interaction
$J_1$ is somewhat stronger than the corresponding intra-bilayer
$c$-axis parameter $J_{c1}$. This suggests the presence of some orbital
polarization in favor of the $x^2-y^2$ state. However, the relatively large
ratio of $\mid J_\text{c1}/J_1 \mid$ as compared to
typical values in the La$_{2-x}$Sr$_x$Mn$_2$O$_7$ system~\cite{Perring,Hirota}
and in copper oxides with bilayer structure~\cite{Tranquada} indicates that
this polarization is far from complete. On the other hand,
$\mid J_\text{c1}/J_1 \mid$ is smaller than the one in isostructural
Ca$_3$Ru$_2$O$_7$~\cite{Nagler}, which presumably reflects the different
electronic structure of the ruthenates where all $4d^4$ valence electrons
reside in the $t_{2g}$ orbitals in a low-spin configuration.


In summary, we have determined the magnetic structure and exchange
interactions of \seven, a clean, stoichiometric compound with a
quasi-two-dimensional spiral state very close to a MIT. The determination of
the spin wave excitations and exchange interactions provides
a quantitative confirmation of a decades-old theoretical
prediction~\cite{deG60} and a firm experimental basis for further experimental
and theoretical work on TMOs near MITs. These include the pseudo-cubic
perovskite Sr(Fe,Co)O$_{3-\delta}$, which exhibits a rich phase diagram as a
function of doping, temperature, and magnetic
field~\cite{Leb04,Adl06,Reehuis2012,Ishiwata2011,Chakraverty}. It was
recently shown that these materials may offer an attractive platform for
exploration of skyrmion physics~\cite{Ishiwata2011,Chakraverty}. Analogous
studies of Sr$_3$Fe$_2$O$_{7-\delta}$ have only recently
begun~\cite{Peets2013}.
The excitation spectrum of a clean quasi-two-dimensional spiral we have
reported here will also serve as a baseline for comparison to neutron
scattering data on magnetically disordered metals near MITs including
isostructural Sr$_3$Ru$_2$O$_7$, whose spin excitations have been discussed in
terms of competing interactions~\cite{Capogna}, and high-temperature
superconducting cuprates, where two-dimensional spiral magnetism is one of the
scenarios that has been invoked to explain the unusual incommensurate spin
excitations observed in the superconducting state~\cite{Lindgard,Sushkov}.

\begin{acknowledgments}
We thank D. Efremov and C. Ulrich for discussions, the members of the Jansen department
and Crystal Growth service group at MPI-FKF for assistance, and the
German Science Foundation (DFG) for financial support under collaborative
grant No.\ SFB/TRR~80.
\end{acknowledgments}

\end{document}